\documentclass[%
reprint,
superscriptaddress,
preprintnumbers,
nofootinbib,
amsmath,amssymb,
aps,
prd,
]{revtex4-1}

\usepackage{hyperref}
\hypersetup{colorlinks=true, citecolor=blue, urlcolor=blue, linkcolor=blue}
\usepackage[T1]{fontenc}
\usepackage{float}      		
\usepackage{dcolumn}    		
\usepackage{bm}         		
\usepackage{graphicx}		
\usepackage{amsmath}		
\usepackage{amssymb}		
\usepackage{aas_macros}
\usepackage{xcolor}
\usepackage{newtxtext,newtxmath}
\usepackage[linesnumbered,ruled,vlined]{algorithm2e}
\usepackage{algpseudocode}
\usepackage{tikz}
\usetikzlibrary{fit,calc}
\usepackage{makecell}

\begin{document}
\title{Tests of Evolving Dark Energy with Geometric Probes of the Late-Time Universe}

\author{Kunhao Zhong}\email{kunhaoz@sas.upenn.edu}
\affiliation{Department of Physics and Astronomy,
University of Pennsylvania, Philadelphia, PA 19104, USA}

\author{Bhuvnesh Jain}
\affiliation{Department of Physics and Astronomy,
University of Pennsylvania, Philadelphia, PA 19104, USA}

\date{\today}

\begin{abstract}
Recent results from the Dark Energy Spectroscopic Instrument (DESI) have shown a strong statistical preference for a time-evolving dark energy model over $\Lambda$CDM when combining BAO, CMB, and supernova (SN) data. We investigate the robustness of this conclusion by isolating geometric information in weak lensing measurements from the DES Year 3 survey and combining it with different datasets. We introduce a hyperparameter, $\Omega_{\rm m}^{\rm growth}$, to decouple the growth contribution from the lensing 2-point correlation and thus bypass the possible effect of the $\sigma_8$ tension in our analysis. We then combine with the late-time geometric probes provided by BAO and SN, along with CMB primary data. The preference for evolving dark energy is consistent with the DESI-DR2 findings: when combining BAO, primary CMB, and weak lensing data, the $w_0w_a$CDM is preferred at about the $3\sigma$ significance. However, when we add SN, the result is sensitive to the choice of data: if we leave out $z<0.1$ SN data in the analysis, as a test of the effect of inhomogeneous calibration, we obtain a statistical significance below $2\sigma$ for time evolving dark energy. Indeed, the high-z only SN data \textbf{lowers} the evidence for evolving dark energy in the data combinations we have examined. This underscores the importance of improved SN samples at low redshift and of alternative data combinations. We show that cosmic shear measurements with LSST Year 1 data will provide comparable power to current SN data. We discuss other low-redshift probes provided by lensing and galaxy clustering to test for evolving dark energy.

\end{abstract}

\maketitle

\section{Introduction}

Recent results from the Dark Energy Spectroscopic Instrument (DESI) and type Ia supernova (SN) challenge the $\Lambda$CDM model, showing a strong statistical preference for time-evolving dark energy~\cite{DESI:2024mwx, DESI:2025zgx, DESI:2025fii, DES:2025bxy}. A joint analysis of baryon acoustic oscillation (BAO), cosmic microwave background (CMB), and SN data prefers the $w_0w_a$CDM model over $\Lambda$CDM by up to $4.2\sigma$. 

The highest significance, $4.2\sigma$, is achieved when combining DESI data with external supernova data from the Dark Energy Survey Year 5 (DES-Y5) and CMB measurements, including the full temperature and polarization acoustic peaks as well as CMB lensing. The DESI Data Release 2~\cite{DESI:2025zgx} (hereafter DESI-DR2) includes a series of robustness tests to verify this potentially groundbreaking result. One such test examines whether the preference for dynamical dark energy is independent of the early-universe CMB constraint on the matter density $\Omega_{\rm m}$. Replacing the CMB data with low-redshift weak lensing measurements from the 3-year Dark Energy Survey (DES-Y3), they found consistent results. However, DES-Y3 data combined with BAO shows only a modest preference for dynamical dark energy ($2.2\sigma$) and does not increase the overall significance when supernova data is included.

In contrast to distance measurements (geometry), constraining structure formation through perturbations (growth) is significantly more challenging. A long-standing debate centers on whether direct probes of structure growth yield a lower clustering amplitude than that predicted by the CMB, a discrepancy known as the $\sigma_8$ tension~\cite{DES:2017qwj, DES:2017myr, DES:2021wwk, Dalal:2023olq, KiDS:2020suj, Stolzner:2025htz}. This difficulty is not unique to weak lensing; it also affects analyses based on redshift-space distortions (RSD), full-shape power spectrum, and other low-redshift measurements. See Ref.~\cite{Abdalla:2022yfr} for a review. Depending on the analysis method and dataset, the inferred lensing amplitude from DESI data varies~\cite{Sailer:2024coh, Chen:2024vvk, DESI:2024jis, DESI:2024hhd, Maus:2025rvz}. There is a mild tension with results from BOSS~\cite{Chen:2024vuf, Chen:2022jzq}. The situation becomes even more complex when the expansion history is allowed to vary within the $w_0w_a$ framework. Nevertheless, with the inclusion of SN data, most analyses still report a consistent preference for the $w_0w_a$ model~\cite{Chen:2024vuf, Karim:2024luk, Maus:2025rvz}. Conversely, whether the dynamical dark energy evolution can help solve the S8 tension is an interesting but still open question~\cite{Heydenreich:2025xim}.

In this work, we aim to simplify the problem by isolating the geometric information from weak lensing probes. By marginalizing over the growth contribution, we can robustly combine weak lensing and BAO data to test whether the preference for dynamical dark energy persists. Our methodology largely follows previous approaches that assess the consistency between geometry and growth~\cite{DES:2020iqt, Ruiz-Zapatero:2021rzl, Zhong:2023how}. When testing the $w_0w_a$ model, we treat the growth information as a nuisance component to marginalize over, rather than attempting to extract precise constraints on the lensing amplitude. By inflating the uncertainty on $\sigma_8$, we test whether the statistical significance of the $w_0w_a$ preference is affected. The main constraining power in this approach arises from the ratios of different two-point statistics.

The structure of this paper is as follows. In Sec.~\ref{sec:methodology} we introduce how we marginalize growth information and other choices different from DESI-DR2, including the effort to better estimate tension. We present results in Sec.~\ref{sec:results} and conclude in Sec.~\ref{sec:conclusion}.

\section{Methodology}\label{sec:methodology}

\subsection{Growth Marginalized Weak Lensing}\label{sec:growth_geometry}

We introduce the growth-geometry split in this section. The analysis mainly follows the previous analysis in Ref~\cite{DES:2020iqt, Zhong:2023how, Ruiz:2014hma, Bernal:2015zom, Andrade:2021njl, 2021A&A...655A..11R}. The goal of this work is not to extract precise information about structure growth, but rather to marginalize over its contribution in order to isolate the geometric component, which can then be combined with BAO and SN data. Thus, we expect the specific implementation choices to have minimal impact on the main results.

To split growth and geometry, we introduce a hyperparameter $\Omega_{\rm m}^{\rm growth}$. Since the late-time linear matter power spectrum can be written as the product of the primordial power spectrum, the transfer function, and the growth function $G_{\rm growth}$, we define the split power spectrum as follows:

\begin{align}\label{def:Pk-slpit}
P_{\rm split}(k , z) &=  P_{\rm geo}^{\rm linear}(k,z) \times  \bigg(\dfrac{G_{\rm growth}^{\rm ODE}(z)}{G_{\rm geo}^{\rm ODE}(z)}\bigg)^2  \times B^{\rm growth}(k, z).
\end{align}

Here, the growth function $G(z)$ describes the growth of structure perturbations. $\times B^{\rm growth}(k, z)$ is the boost factor that corrects for non-linear effects. 

$G(z)$ is a function of parameters $\Omega_{\rm m}$, $w_0$, and $w_a$:

\begin{equation}\label{eq:G_eq}
G^{\prime \prime}+\left(4+\frac{H^{\prime}}{H}\right) G^{\prime}+\left[3+\frac{H^{\prime}}{H}-\frac{3}{2} \frac{\Omega_{\mathrm{m}}^0 a^{-3}}{E^2(a)}\left(1-f_v\right)\right] G=0 ,
\end{equation}
where in $w_0w_a$CDM:
\begin{equation}
\begin{aligned}\label{eq:E2_eq}
    E^2(a) & \equiv H(a)^2/H_0^2 \\
    & = \Omega_{\mathrm{m}}^0 a^{-3}+\Omega_{\mathrm{r}}^0 a^{-4}+\Omega_{\mathrm{\Lambda}}^0 a^{-3\left(1+w_0+w_{\mathrm{a}}\right)} e^{-3 w_{\mathrm{a}}(1-a)} .
\end{aligned}
\end{equation}
We assume a flat universe $\Omega_{\mathrm{\Lambda}}^0 = 1 - \Omega_{\mathrm{m}}^0$. Neutrino effect is accounted for through the fractional mass term $f_\nu = \Omega_\nu/\Omega_\mathrm{m}$. Prime in eq.~\ref{eq:G_eq} denotes derivative with respect to $\log a$.

\begin{figure}[t]
\includegraphics[width=\columnwidth]{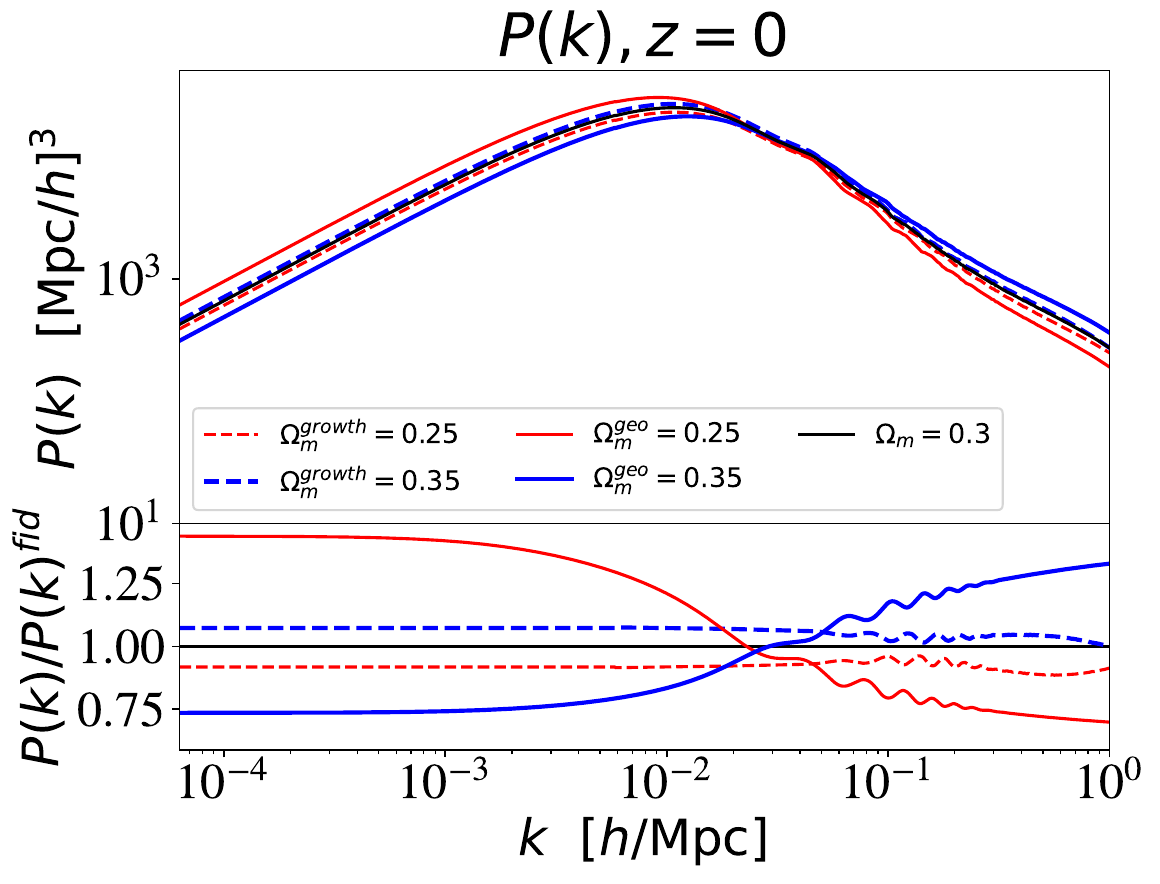}
\caption{The effect of $\Omega_{\rm m}^{\rm growth}$ on the power spectrum. The effect of the growth parameter is much weaker than the geometry parameter, and it primarily changes the amplitude of the power spectrum. Thus marginalizing over $\Omega_{\rm m}^{\rm growth}$ leaves cosmic shear insensitive to the amplitude of matter fluctuations. The oscillatory features at high $k$ are due to non-linear matter evolution. This figure is reproduced from Ref.~\cite{Zhong:2023how}}
\label{fig:pksplit}
\end{figure}

Eq.~\ref{eq:G_eq} is a function of background parameters. Thus, by redefining the power spectrum as eq.~\ref{def:Pk-slpit}, we separated the geometry part of the power spectrum for the combination of other geometry probes, BAO and SN. The split form of the power spectrum has the growth parameter $\Omega_{\rm m}^{\rm growth}$ controlling the evolution of the perturbation while having the normal parameters controlling the shape. This is also the equivalent to marginalizing out the amplitude of $P(k)$, as illustrated in Fig.~\ref{fig:pksplit}. The advantage of introducing this hyperparameter is that it can also be used in other growth-related calculations, such as the non-linear power spectrum boost $B^{\rm growth}(k, z)$ and intrinsic alignment of weak lensing fields.

In a model with dynamical dark energy, such as $w_0w_a$CDM, the growth function is also a function of $w_0$ and $w_a$. One could introduce another set of growth parameters, such as $w^{\rm growth}$ used in Ref.~\cite{DES:2020iqt}. However, this will degrade the constraining power by a lot.  By not separating the $w_0$ and $w_a$ parameter, we assume that the perturbation equation follows the same $w_0$ and $w_a$. Again, the main purpose of this work is not to obtain accurate growth information, and a single parameter $\Omega_{\rm m}^{\rm growth}$ is flexible enough to marginalize out the amplitude of the power spectrum, thus alleviating the concerns about $\sigma_8$ tension. We show how $\Omega_{\rm m}^{\rm growth}$ changes only the amplitude of the power spectrum while $\Omega_{\rm m}^{\rm geo}$ changes the shape of the power spectrum in Fig.~\ref{def:Pk-slpit}.

The oscillatory feature of the $P(k)$ change with the growth factor in Fig.~\ref{fig:pksplit} is due to the non-linear boost, since we split the power spectrum based on the linear $P(k)$. There is no a priori reason to rescale linear $P(k)$ or non-linear $P(k)$ in this case. How to model the non-linear part in the presence of phantom crossing is still an open question, see Ref.~\cite{Dakin:2019vnj, Euclid:2020rfv} for details. We use $\Omega_{\rm m}^{\rm growth}$ in the non-linear emulator to better mitigate this uncertainty. 

For other systematics in weak lensing probes, we follow Ref.~\cite{DES:2021wwk}. The only difference is we use the non-lionear alignment (NLA) model~\cite{Hirata:2004gc} for intrinsic alignment, which is a simplified version of Tidal Alignment and Tidal Torquing (TATT)~\cite{Blazek:2017wbz} used in Ref.~\cite{DES:2021wwk}. This largely reduce the computational requirement, and it was tested in Ref.~\cite{Zhong:2023how} that it does not bias the results when growth is marginalized. As a result, we expect the constraints on $w_0$–$w_a$ to remain robust. However, as noted in recent studies~\cite{Chen:2023yyb, Vlah:2019byq}, both NLA and TATT assume a smooth redshift evolution of alignment parameters—an assumption that, in principle, should be relaxed.  The use of geometry–growth separation partially mitigates this issue.

We can then jointly fit the weak lensing data, BAO from DESI, and SN. The BAO data used in this work is DESI-DR2, and the SN data used is DES-Y5. We will first check whether the conclusions in DESI-DR2 about combining weak lensing changes after the growth information is marginalized. Then, we further explore different combinations of weak lensing data. The details of validation process is discussed in Sec.~\ref{sec:other_diff}. 

We perform the joint analysis using \texttt{Cocoa}, the \texttt{Cobaya}\cite{Torrado:2020dgo}-\texttt{Cosmolike}~\cite{Krause:2016jvl} Architecture. The BAO and SN likelihood are implemented in \texttt{Cobaya} and weak lensing data vector and analytical covariances are calculated with \texttt{Cosmolike}. We use metropolis hasting MCMC sampler, and the Gelman-Rubin criteria~\cite{10.1214/ss/1177011136} $R-1 < 0.03$ for chain convergence.

\subsection{Tension Metric}\label{sec:tension_metric}

To quantify the statistical significance, we calculate the chi-square difference of the two maximum a posteriori (MAP) points from $\Lambda$CDM and $w_0w_a$CDM model, $\Delta \chi_{\mathrm{MAP}}^2$. Following DESI-DR2 approach, we convert the $\Delta \chi_{\mathrm{MAP}}^2$ to the effective marginalized $N \sigma$ significance via:

\begin{equation}
\mathrm{CDF}_{\chi^2}\left(\Delta \chi_{\mathrm{MAP}}^2 , 2 \ \mathrm{dof}\right)=\frac{1}{\sqrt{2 \pi}} \int_{-N}^N e^{-t^2 / 2} d t ,
\end{equation}
where the right hand side is the CDF of chi2 distribution of 2 degrees of freedom, evaluated at $\Delta \chi_{\mathrm{MAP}}^2$. Equivalently:

\begin{equation}
N = \sqrt{2} \cdot \operatorname{erf}^{-1} \left( \operatorname{CDF}_{\chi^2}(\Delta \chi^2_{\text{MAP}} , 2 \ \mathrm{dof}) \right) .
\end{equation}

For example, with two extra degrees of freedom $w_0$ and $w_a$, a $\Delta \chi^2$ improvement of 12 translates to roughly $3\sigma$ significance of ruling out $\Lambda$CDM. More sufisticated tension metrics relaxing different assumptions and considering projection effects can be used to assess the tension between different models or different probes~\cite{Raveri:2018wln, Park:2019tyw, Adhikari_2019, DES:2020hen, Raveri:2021wfz, Saraivanov:2024soy, Raveri:2024dph}. See Ref.~\cite{Efstathiou:2025tie} for potential issues of using this simple tension metric. In this study, we only show $\Delta \chi_{\mathrm{MAP}}^2$ and significance to be directly compared to DESI-DR2.

The likelihood of DES-Y3, however, is very noisy. As a result, the minimizer may not be able to find the global maximum of the posterior. We attempted to obtain the MAP using the \texttt{Py-BOBYQA}~\cite{2018arXiv180400154C} minimizer implemented in \texttt{Cobaya}\footnote{DESI-DR2 uses minimizer \texttt{iminuit}~\cite{dembinski_2025_15157028}. We tested that both minimizer gives the same significance for BAO+SN chains.}. However, we find that the maximized log posterior does not change much from the maximum of the sampled chain. For some of the nuisance parameters, the value does not move from the initial points. This indeed indicates that the minimizer does not work well.

\begin{figure}[t]
\includegraphics[width=1\columnwidth]{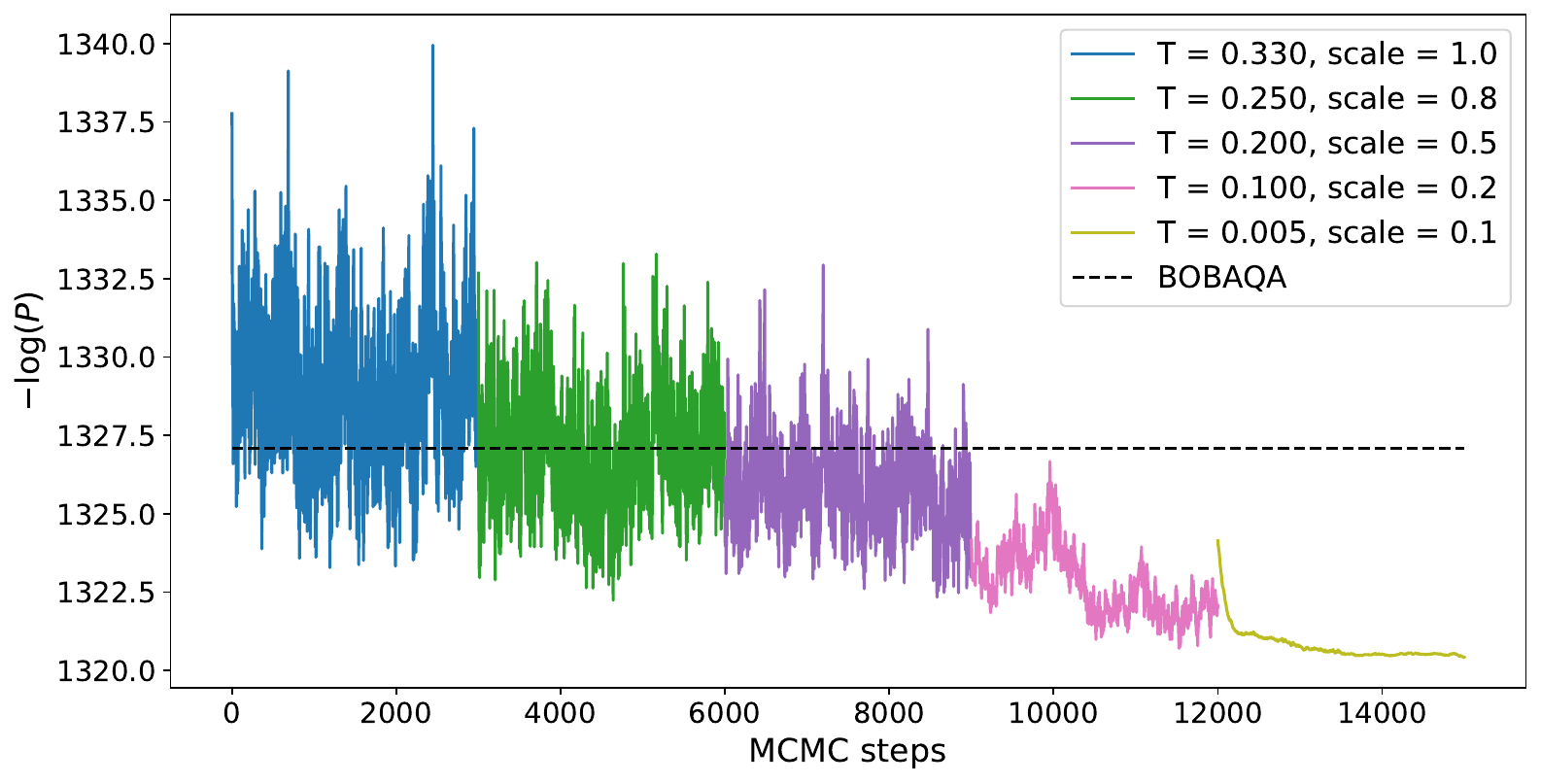}
\caption{The log posterior as a function of time steps, in the tempered MCMC approach. At each stage, the temperature and scale factor are decreased, forcing the walkers to search optimal values within a confined spaced.}
\label{fig:t_mcmc}
\end{figure}

To address this issue, we adopted the tempered MCMC approach, also called simulated annealing, similar to Ref.~\cite{Karwal:2024qpt}. The main idea is to continue running a converged MCMC chain with an annealing temperature. The temperature of the MCMC exponentially suppresses the likelihood ratio $\left(\mathcal{L}_{\text {new }} / \mathcal{L}_{\text {old }}\right)^{1 / T}$. As a result, the sampler will reject most of the points and only sample in a confined place near the best-fit point so far. As shown in Ref.~\cite{Karwal:2024qpt}, by gradually decrease the temperature and step size, this stochastic approach is able to find the minimum much more efficient than a non-gradient-based method. We test the same method and show the results in Fig.~\ref{fig:t_mcmc}. The MAP value obtained through tempered MCMC is indeed about $\Delta \chi^2 \approx 7$ lower than that from the standard minimizer.

However, we find that using the tempered MCMC does not change the relative difference between the two models, $\Lambda$CDM and $w_0w_a$CDM. The MAP value decreases by nearly the same amount in both models compared to the minimizer-based approach. Although the impact on the final significance and tension metrics is negligible, we emphasize the importance of a robust MAP-finding algorithm when working with noisy datasets such as weak lensing.

\subsection{Other Analysis Differences}\label{sec:other_diff}

\begin{figure}[t]
\includegraphics[width=1.0\columnwidth]{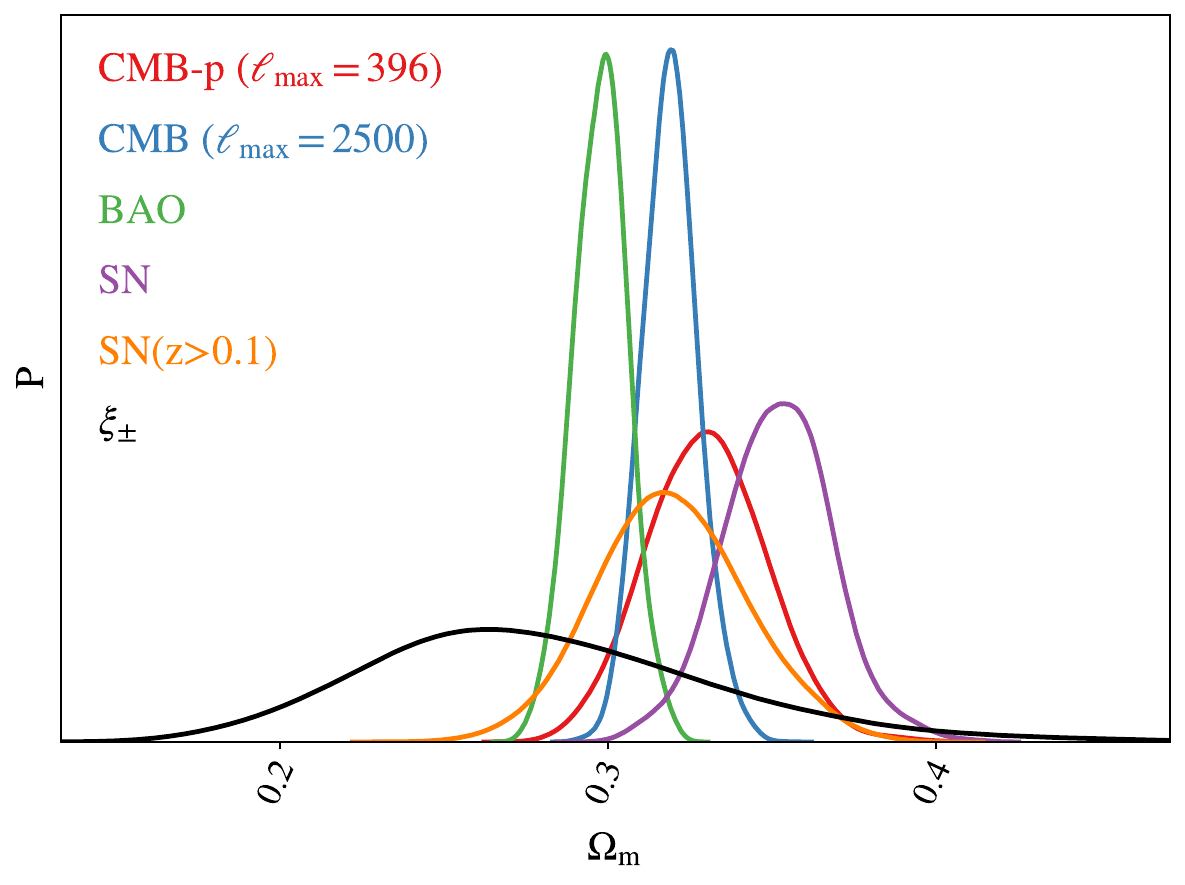}
\caption{1D marginalized posterior for $\Omega_{\rm m} $ in the $\Lambda$CDM model. The main source of preference for evolving dark energy is the tension between  BAO and CMB. The tension is even higher between  BAO and DES-Y5 SN. However,  if low-z SN are excluded, the tension gets much weaker (compare the purple and orange posteriors). Indeed, we find that $z>0.1$ SN lower the evidence for evolving dark energy in all data combinations we have studied. Cosmic shear also prefers a low $\Omega_{\rm m} $, although the uncertainty is relatively large.  We have shown two choices of the CMB posterior: the one with higher 
$\ell_{max}$ is shown for reference -- see discussion in Sec.~\ref{sec:other_diff}. 
}
\label{fig:omegam}
\end{figure}


Apart from the effort to isolate geometry information from DES-Y3, we further validate other analysis choices in DESI-DR2. We summarize them below.\\ \\

\textbf{Cosmic Shear $\xi_\pm$ rather than full 3x2pt.} The DES-Y3 key product contains auto-correlation of galaxy shapes (cosmic shear $\xi_\pm$), auto-correlation of galaxy positions (galaxy clustering), and cross-correlation between them (galaxy-galaxy lensing). The combination of them all is often referred to as 3x2pt. Using the cross correlations including galaxy positions increase the constraining power significantly; the length of the data vector doubles. However, this comes at the cost of increased model complexity, particularly due to the need for galaxy bias modeling. This issue is especially relevant because one of the lens samples in DES-Y3 shows internal inconsistencies; see Ref.~\cite{DES:2021wwk, DES:2021zxv} for more detailed discussion. In this work, we retain the cosmic shear data to avoid unknown systematics~\footnote{An updated $\xi_\pm$ measurement is used in Ref.~\cite{DES:2024xvm}. In this work, we tested that the updated and the public data vector give identical results. }. A future work will extend the work in this paper to 3x2pt to gain more statistical significance. \\ \\
\textbf{Combination of posteriors.} In DESI-DR2, the joint posteriors are not sampled directly. They use \texttt{CombineHarvesterFlow}~\cite{Taylor:2024eqc} to learn the joint posterior using normalizing flow (NF). There are also other similar effort with different implementation such as Ref.~\cite{Raveri:2024dph}. However, as noted in \texttt{CombineHarvesterFlow}~\cite{Taylor:2024eqc}, this method could fail in two cases. The first one is when two experiments are in tension under the same model, which is the main concern that this work is trying to address. The second one is when combining two experiment that has very different constraining powers, which is likely when the CMB and SN dataset are present. For more details, see Ref.~\cite{Taylor:2024eqc}. A practical limitation of NF-based methods is the lack of a well-established convergence criterion to assess the accuracy of the learned transformation. Thanks to the \texttt{Cocoa} pipeline, we can directly sample DESI-BAO, SN, and DES-Y3 likelihood together. Validation the accuracy of the NF method is beyond the scope of this paper. But we stress that the direct joint sampling should be more robust.\\ \\

\textbf{Primary CMB priors. } DESI-DR2  choose a three-parameter multivariate Gaussian likelihood for the CMB: the parameters are $(\theta_, w_b, w_{bc})$~\footnote{Equation A1-A2 of Ref.~\cite{DESI:2025zgx}.}. They are constructed such that late-time effects—namely, the late integrated Sachs–Wolfe (ISW) effect and CMB lensing—are marginalized out. Conceptually, this is aligned with the goal of this paper: isolating geometric information from late-time perturbations. 

However, our goals in detail are different: in order to use the weak lensing likelihood, we require some input on the primordial perturbations, particularly $A_s$ and $n_s$, while marginalizing over the late-time evolution using $\Omega_{\rm m}^{\rm growth}$. Moreover as shown by \cite{DESI:2025gwf} (see e.g Fig. 1), different choices of CMB data show significant variations in the tension with BAO (within LCDM) despite small changes for the posteriors. Note that CMB lensing contributed significantly to the $\Delta \chi^2$ in DESI-DR2. To avoid this uncertainty and keep the focus on late time probes, we use minimal low-$\ell$ information from the CMB via the "primary CMB" (CMB-p) dataset. This includes the Planck 2018 high-$\ell$ TT, TE, and EE spectra, truncated after the first acoustic peak ($35 < \ell < 396$), and the low-$\ell$ EE polarization data ($\ell < 30$). This configuration removes sensitivity to the late-time ISW and lensing contributions in TTTEEE, and we explicitly exclude the CMB lensing likelihood. Another choice that has similar goals would be including a broader $n_s$ prior, but we consider the priomary CMB as a more direct and natural choice. The CMB-p prior on $(\theta_, w_b, w_{bc})$ is significantly broader than the Gaussian prior used in DESI-DR2, but the constraints on dark energy parameters are very similar for either choice of CMB data as we shown in Appendix~\ref{sec:cmb_likelihoods}. 
See also Ref.~\cite{DESI:2025gwf} on the discussion about the impact of the latest ACT-DR6 measurements.
\\ \\
\textbf{Type Ia supernova data.} The SN dataset used in this work is DES-Y5~\cite{DES:2024jxu}, which includes 1,635 photometrically classified supernovae at $z > 0.1$ and 194 low-redshift supernovae from historical surveys. Since the SN dataset is one of the primary drivers of the $w_0w_a$ preference, there has been some debate about the calibration process, particularly for the low-redshift component~\cite{Efstathiou:2024xcq, DES:2025tir}. One option, as adopted in DESI-DR2, is to test for calibration issues by imposing a cut at $z = 0.1$. We also investigate the subset of high redshift only SN, dubbed as SN (z>0.1). Note that in this work, we put a direct mask on the SN data and covariance matrices to impose the redshift cut, whereas in principle, a refit to some calibration parameters is needed. However, as shown in Appendix C of Ref.~\cite{DES:2024jxu}, this effect is small. Aside from this test, we use the full DES-Y5 sample without such a cut and refer to it simply as "SN" throughout the paper. The other two SN samples used in the DESI-DR2 analysis show a weaker preference for evolving dark energy. We use only the DES-Y5 sample as our focus is on weak lensing data, and the DES sample illustrates the strongest variation evidence when adding SN. 

\section{Results}\label{sec:results}

\begin{figure*}[t]
\includegraphics[width=1.0\columnwidth]{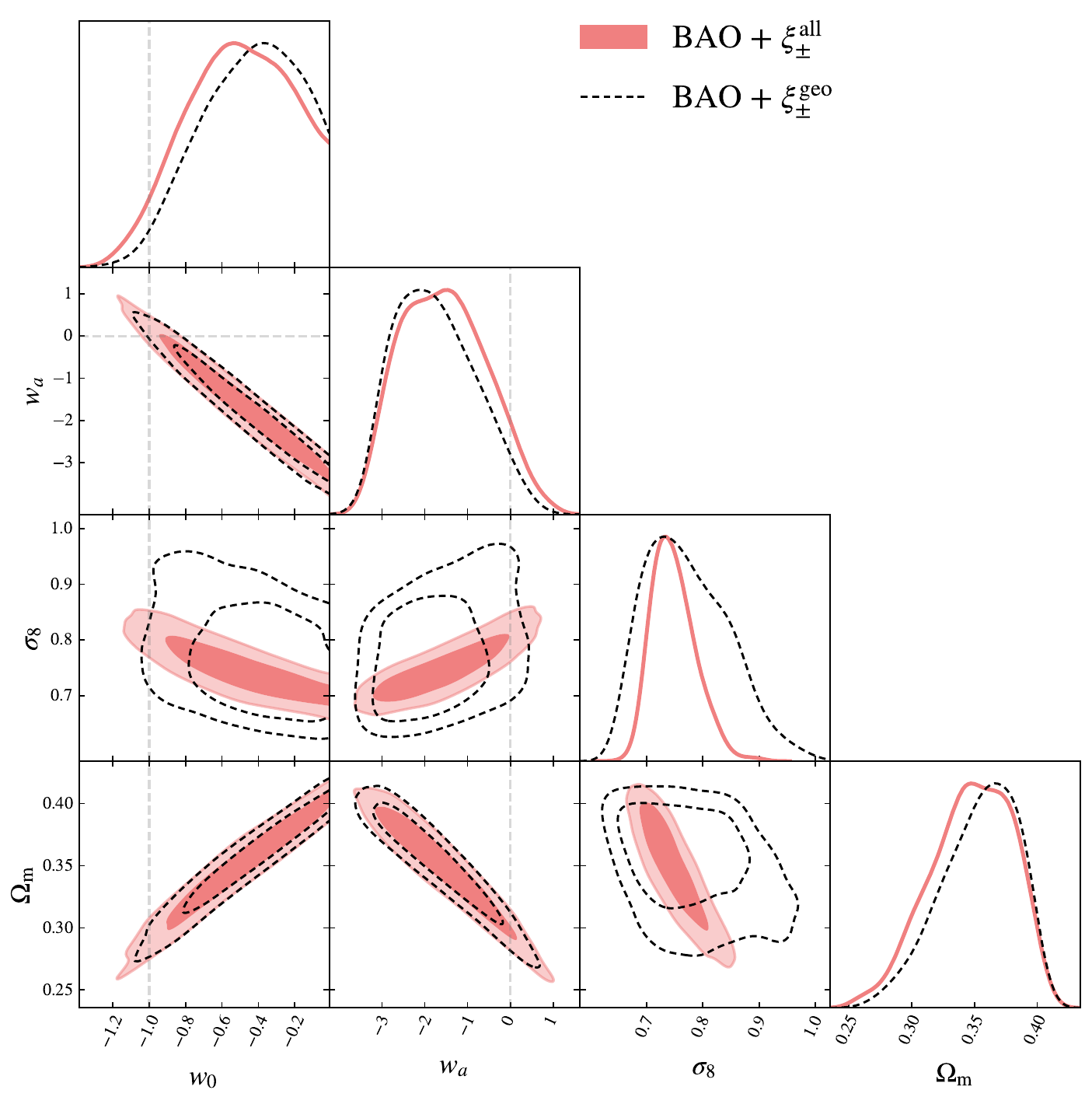}
\includegraphics[width=1.0\columnwidth]{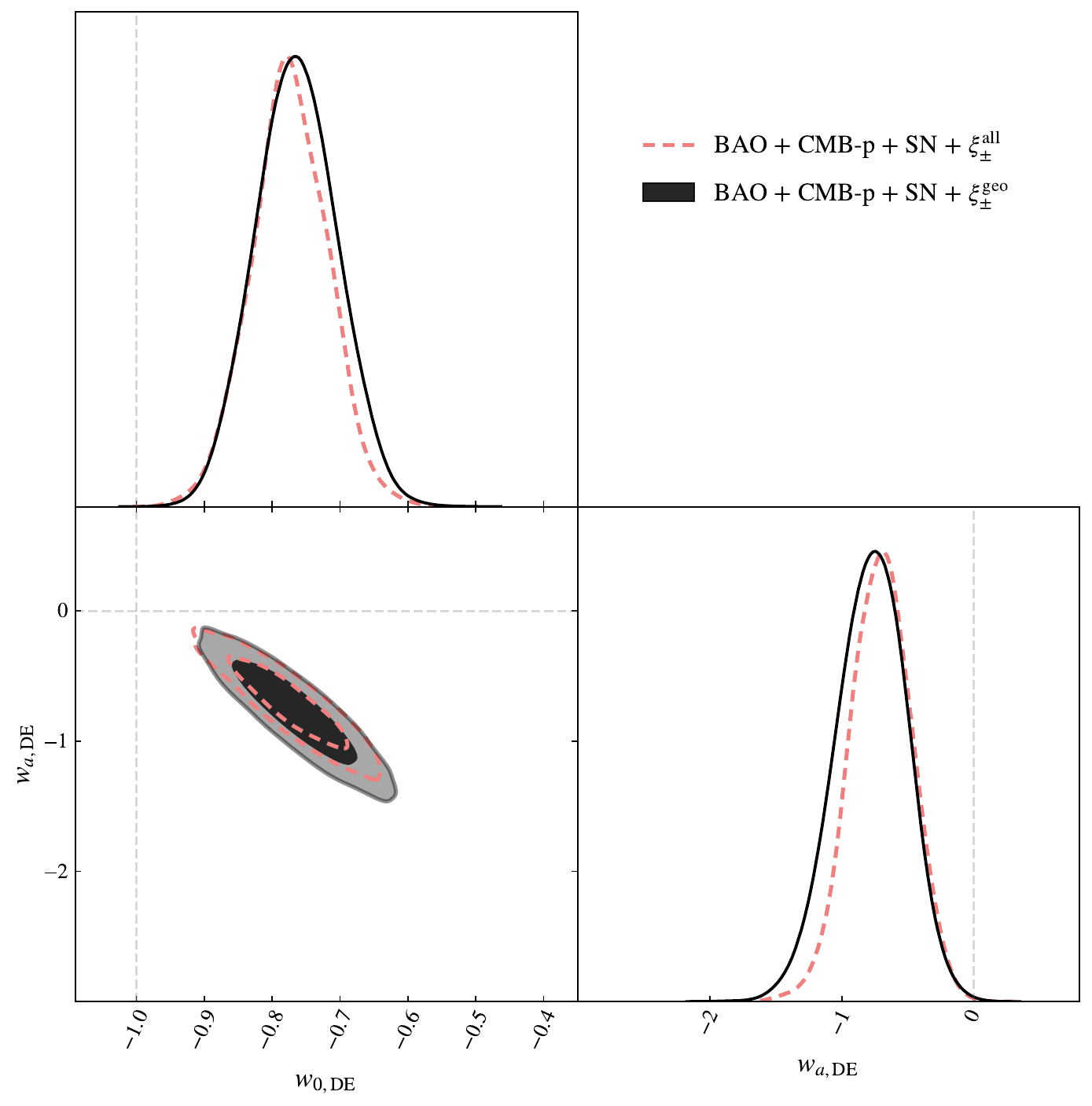}
\caption{\textit{Left Panel:} Posterior in the $w_0-w_a$ plane for the combination of BAO and cosmic shear, with and without the growth–geometry split ($\xi_\pm^{\rm geo}$ and $\xi_\pm^{\rm all}$). Introducing the additional hyperparameter $\Omega_{\rm m}^{\rm growth}$ leads to an unconstrained posterior on $\sigma_8$, as expected (it is prior dominated). However, the constraints on $w_0$ and $w_a$ remain almost unchanged, which shows that the geometric part of cosmic shear dominates information on dark energy. \textit{Right Panel:} Posterior distributions for the combination of BAO, primary CMB (CMB-p), SN, and cosmic shear. The geometry isolation only slightly degrades the constraining power. 
}
\label{fig:gg_comp}
\end{figure*}

\begin{figure*}[t]
\includegraphics[width=\columnwidth]{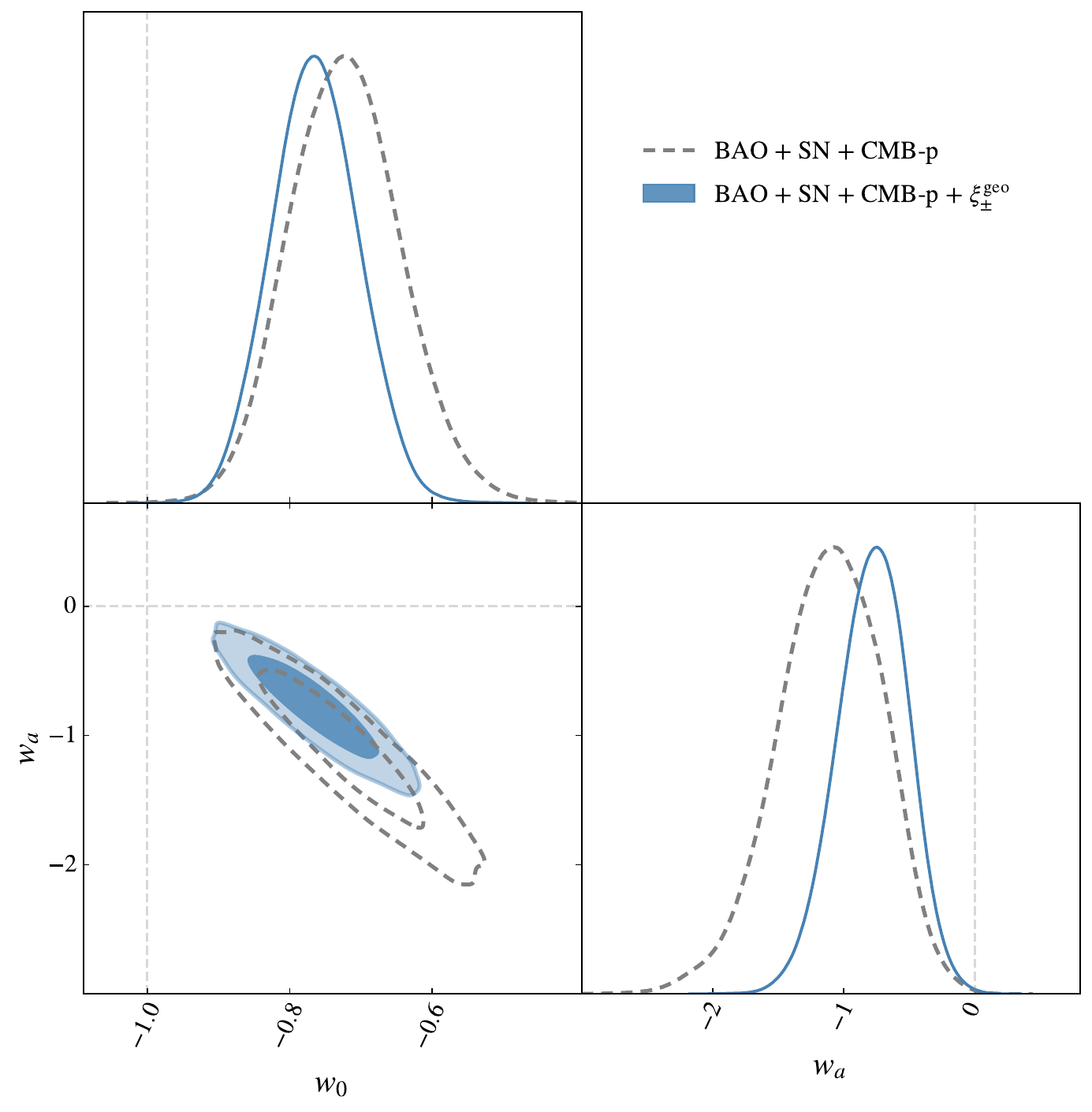}
\includegraphics[width=\columnwidth]{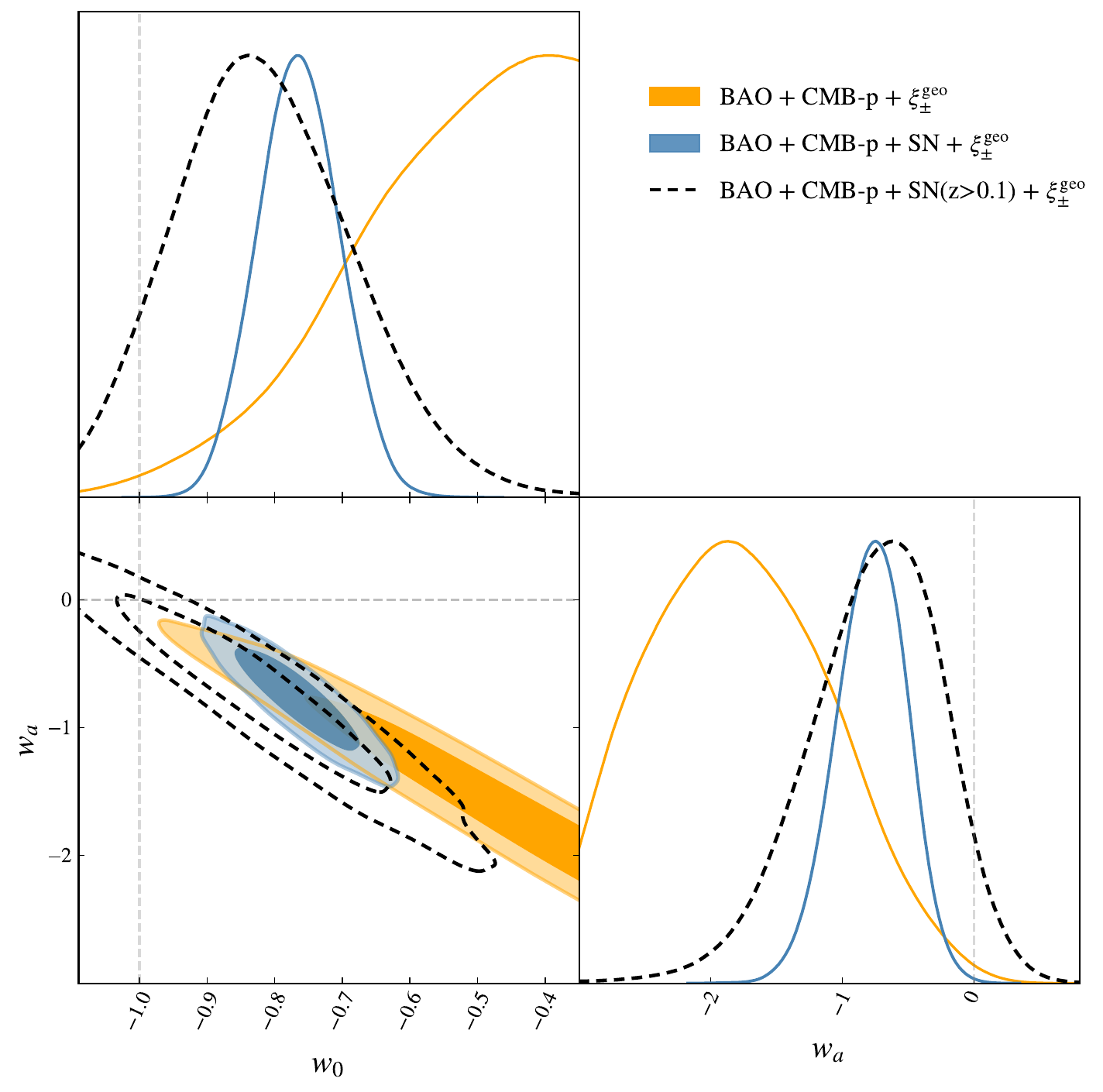}
\caption{\textit{Left Panel:} Posterior in the $w_0-w_a$ plane for the combination  BAO, SN, CMB-p and $\xi_{\pm}^{\rm geo}$. The addition of cosmic shear in this case improves constraints on $w_0$ and $w_a$  and increases the statistical significance in favor of evolving dark energy by $0.3\sigma$. \textit{Right Panel:} Posterior with different choices of SN data: no SN (orange, filled contours); full DES SN sample (blue filled contours) and $z>0.1$ SN sample (dashed contours). The latter case shows consistency with LCDM at the $2-\sigma$ level, showing the strong dependence of the evolving dark energy result on the low $z$ SN. 
}
\label{fig:posterior_main}
\end{figure*}

\begin{figure}[t]
\includegraphics[width=1.0\columnwidth]{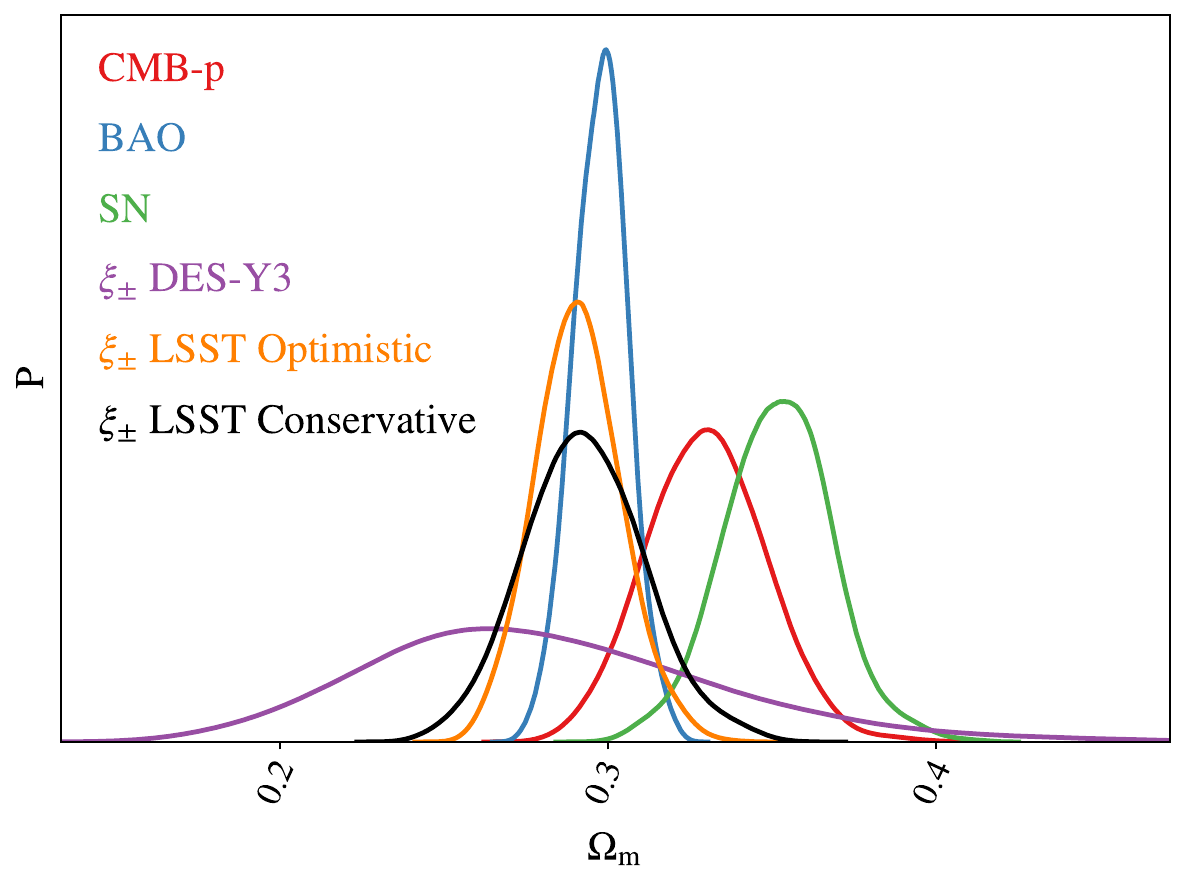}
\caption{Forecast of the LSST-Y1 cosmic shear constraint on $\Omega_{\rm m}$. The "Optimistic" case follows the uncertainty of the photo-z in Ref.~\cite{Eifler:2020vvg} while the "Conservative" uses the same prior as DES-Y3. The other choices follow Ref.~\cite{Eifler:2020vvg}. The real constraining power will likely lie in between, which is  a significant boost and comparable to SN and CMB-P.}
\label{fig:lsst_forcast}
\end{figure}

\begin{table}[t]\label{tab:tension}
\renewcommand{\arraystretch}{1.3}
\centering
\begin{tabular}{lcccc}
\hline\hline
Datasets & $\Delta\chi^2_{\mathrm{MAP}}$ & Significance \\
\hline
BAO + CMB-p & $-6.9$ & $2.1\sigma$  \\
BAO + $(\theta_*, w_{b/bc})_{\mathrm{CMB}}$ & $-8.0$ & $2.4\sigma$  \\
BAO + SN & $-13.6$ & $3.3\sigma$  \\
BAO + CMB-p + SN  &  -15.5      & $3.5\sigma$ \\ 
BAO + $(\theta_*, w_{b/bc})_{\mathrm{CMB}}$ + SN &  -14.2      & $3.4\sigma$ \\
\hline
BAO + CMB-p  + $\xi_{\pm}^{\rm geo}$ &  -8.4      &  $2.4\sigma$ \\
BAO + $(\theta_*, w_{b/bc})_{\mathrm{CMB}}$ + $\xi_{\pm}^{\rm geo}$ &  -8.7      & $2.5\sigma$ \\
BAO  + SN + $\xi_{\pm}^{\rm geo}$ &  -13.6      & $3.3\sigma$ \\
BAO + CMB-p + SN + $\xi_{\pm}^{\rm geo}$ &  -17.2      & $3.8\sigma$ \\
BAO + $(\theta_*, w_{b/bc})_{\mathrm{CMB}}$ + SN + $\xi_{\pm}^{\rm geo}$ &  -18.8      & $3.9\sigma$ \\
\hline
\end{tabular}
\caption{Different data combinations and the corresponding $\Delta\chi^2_{\mathrm{MAP}}$ and statistical significance of departures from $\Lambda$CDM. We checked that the numbers for BAO+SN match those in Ref.~\cite{DESI:2025zgx}. Here, CMB-p refers to primary CMB as discussed in Sec.~\ref{sec:other_diff}, and $(\theta_*, w_{b/bc})_{\mathrm{CMB}}$ is the three-parameter compressed prior.
}
\end{table}

Before showing the posterior distribution of combined probes, we show the 1D marginalized posterior of $\Omega_{\rm m}$ in Fig.~\ref{fig:omegam}, assuming $\Lambda$CDM background. Despite the complications in high-dimensional space and different degeneracies from different probes, this 1D distribution is very informative. As shown in Ref.~\cite{Tang:2024lmo}, the $w_0w_a$ tension can be roughly translate to an $\Omega_{\rm m}$ tension. The discrepancy of $\Omega_{\rm m}$ value between BAO and SN can only be solved with a strong preference for evolving dark energy. The story is similar for other probes. For example, the shift to a smaller $\Omega_{\rm m}$ for high-z SN means it is more consistent with BAO and $\Lambda$CDM. It is interesting that in DES-Y3, the posterior of $\Omega_{\rm m}$ shifts when galaxy clustering and galaxy-galaxy lensing are included.

We first test whether marginalizing the growth information affects the posterior distribution. As shown in Fig.~\ref{fig:gg_comp}, the posterior on $w_0$–$w_a$ remains nearly unchanged with or without the growth–geometry split. However, marginalizing over growth leads to a significantly broader posterior on $\sigma_8$ in the $\xi_\pm$ analysis, which provides a more conservative and robust framework for combining with CMB data. This marginalized treatment ensures that the posterior on late-time geometric information is not biased by projection effects, particularly in the presence of $\sigma_8$ tension.

As noted in DESI-DR2, the most constraining dataset combination is DESI, DES Y5 SN, and the full CMB. The marginalized posteriors for this combination are:
\begin{equation}
\left.
\begin{aligned}
w_0 &= -0.752 \pm 0.057 \\
w_a &= -0.86^{+0.23}_{-0.20}
\end{aligned}
\right\} \quad \text{BAO + CMB (full+lensing) + SN,}
\end{equation}
When we use primary CMB instead, we get slightly degraded constraints:
\begin{equation}
\left.
\begin{aligned}
w_0 &= -0.721^{+0.082}_{-0.072} \\
w_a &= -1.11^{+0.36}_{-0.45}
\end{aligned}
\right\} \quad \text{BAO + CMB-p + SN}
\end{equation}

However, when cosmic shear information is added, we obtain stronger constraints:
\begin{equation}
\left.
\begin{aligned}
w_0 &= -0.764^{+0.061}_{-0.060} \\
w_a &= -0.781^{+0.25}_{-0.29}
\end{aligned}
\right\} \quad \text{BAO + CMB-p + SN + $\xi_\pm^{\rm geo}$,}
\end{equation}

Note that this is obtained without CMB lensing and only uses the first peak of the  CMB power spectrum, which is measured in high precision. When we add more acoustic peaks to $\ell_{\rm max}=900$, the error bar is tighter. However, the significance remains the same at 3.8 sigma.

The statistical significances of departures from LCDM are summarized in Table~\ref{tab:tension}. The addition of $\xi_\pm^{\rm geo}$ increases the significance by approximately $0.3\sigma$ when combined with either BAO+CMB-p+SN or BAO+CMB-p, but has negligible impact on the BAO+SN combination. This is mainly because  CMB-p provides a meaningful prior on the power spectrum, particularly on $n_s$, which makes $\xi_\pm^{\rm geo}$ more constraining for dark energy. Overall, these results reinforce the main conclusion of DESI-DR2: $\Lambda$CDM is  disfavored in favor of evolving dark energy models with phantom crossing. 

However, the story changes when we apply a redshift cut at $z > 0.1$ in the SN sample. There is extensive discussion in the literature about the inhomogeneous SN sample at $z<0.1$, which makes calibration more challenging~\cite{Popovic:2025glk}. We find, across all tested data combinations, no case where the tension exceeds $\approx 2\sigma$ when we only use high redshift SN with $z>0.1$. The constraints are: 
\begin{equation}
\left.
\begin{aligned}
w_0 &= -0.81\pm0.14 \\
w_a &= -0.781\pm0.54
\end{aligned}
\right\} \quad \text{BAO + CMB-p + SN($z>0.1$) + $\xi_\pm^{\rm geo}$,}
\end{equation}

This finding is consistent with the DESI-DR2 findings, which used CMB lensing instead of galaxy lensing and underscores the need for improved SN data at low redshift and for continued studies with alternative data combinations to test for evolving dark energy. Upcoming SN samples from the ongoing DEBASS project ~\cite{Acevedo:2025fpo, Sherman:2025rgl} and in the future from LSST~\cite{LSSTDarkEnergyScience:2022oih} and ZTF~\cite{Rigault:2024kzb} are expected to provide more homogeneous low-$z$ calibration, helping to reduce systematic uncertainties in future analyses.

As discussed in Sec.~\ref{sec:other_diff}, we restrict our analysis to the cosmic shear component of the weak lensing data rather than the 3x2pt data vector. We tested a synthetic  3x2pt data vector to assess potential improvements. We find that the gains are primarily in constraining growth-related parameters, while the $w_0$–$w_a$ constraints remain nearly unchanged compared to the results presented here. This may be due in part to the fiducial DES-Y3 analysis choice of excluding the two highest redshift bins for the lens sample. These high-$z$ bins, probing the matter-dominated era, could in principle contribute more geometric constraining power.

\section{Conclusion}\label{sec:conclusion}

We revisited the recent DESI-DR2 results that show a strong statistical preference for a time-evolving dark energy model over a cosmological constant, with a focus on assessing the robustness of this conclusion using different low-redshift probes. Specifically, we isolate the geometric information from weak lensing by marginalizing over growth using a hyperparameter $\Omega_{\rm m}^{\rm growth}$. This approach allows weak lensing data from the DES-Y3 survey to be combined more robustly with BAO, SN, and primary CMB measurements, excluding CMB lensing and ISW contributions, and excluding potential biases from the $\sigma_8$ tension.

Our results show that marginalizing growth does not weaken the preference for the $w_0w_a$CDM model. It improves robustness by eliminating spurious constraints arising from projection effects and growth-dependent systematics. We validated our analysis choices, including the use of primary CMB information (CMB-p), exclusion of potentially inconsistent galaxy lensing samples, and tempered MCMC methods for tension calculation. The combination of DESI-DR2 BAO, DES-Y5 SN, CMB-p, and DES-Y3 $\xi_\pm^{\rm geo}$ gives a strong preference for dynamical dark energy, even without using the full CMB lensing likelihood.

We also examine the possible discrepancy between low-$z$ and high-$z$ SN data in the DES-Y5 sample. We find that using $z>0.1$ SN \textbf{lowers} the evidence for evolving dark energy in all the data combinations we have examined, contrary to the findings from the full SN sample. In summary we find that if the $z<0.1$ SN are excluded, the evidence for evolving dark energy is typically about $2\sigma$ depending on the datasets combined. Thus, it is crucial to obtain reliable low-redshift SN to increase confidence in the significance of the breakdown of LCDM. See also Ref.~\cite{Efstathiou:2024xcq, Cortes:2025joz} for potential issues in including SN in $w_0w_a$ analysis. In the meantime, progress in creating a consistent SN dataset as the low-z anchor is ongoing -- in particular, the DEBASS survey, which has published promising prelininary results~\cite{Acevedo:2025fpo, Sherman:2025rgl}. 

Independent low-redshift probes will also be critical for future constraints on evolving dark energy. Galaxy clustering, lensing and cross-correlations with bigger samples are one avenue. Kinematic lensing~\cite{Xu:2024qyk} is promising as an alternative as it significantly decreases the shape noise contribution and boosts sensitivity to the low-$z$ lensing signal. 

While the statistical contribution of DES Y3 cosmic shear is weaker than that of current BAO and SN measurements, upcoming surveys such as Euclid~\cite{laureijs2011euclid}, LSST~\cite{LSST:2008ijt}, and the Roman Space Telescope~\cite{Dore:2019pld} will provide significantly more powerful constraints. We forecast the potential contribution from the LSST-Y1 survey, which is scheduled to begin by end of 2025. A simple metric is the expected marginalized constraint on $\Omega_{\rm m}$, which we find to be significantly tighter than that from DES, as shown for two scenarios in Fig. \ref{fig:lsst_forcast}. The optimistic and conservative cases differ in their assumptions about photometric redshift (photo-$z$) uncertainties: the optimistic case follows the specifications in Ref.\cite{Eifler:2020vvg}, while the conservative case adopts DES-like values. Even under conservative assumptions, the improvement in statistical power due to increased galaxy number density and reduced uncertainties suggests that LSST-Y1 will be a highly competitive probe in the very near future.

There exist other ways, potentially more constraining, to isolate geometric information. While we limit our primary analysis to the cosmic shear component of DES-Y3, including full 3x2pt information in future analyses is valuable provided the measurements and modeling of galaxy clustering and cross-correlation are robust. One could also take lens samples at relevant redshifts and cross correlates with the CMB. Several galaxy $\times$  CMB analyses have been carried out ~\cite{DES:2022xxr, Xu:2023qmp, Yao:2023cxv, Qu:2024sfu}, but the constraining power for $w_0-w_a$ has not been fully explored. These analyses require using galaxy clustering measurements to  marginalize over galaxy bias. Cross-correlating galaxy lensing shear with CMB lensing avoids the need to marginalize over galaxy bias. Another possible approach is the shear ratio method~\cite{Jain:2003tba, DES:2021jzg, Emas:2024kwd}, which takes the ratio of galaxy–galaxy lensing signals using the same lens sample, thereby canceling the dependence on the lensing amplitude.

Future extensions of this work could incorporate various lens samples, shear ratio measurements, cross-correlations with CMB lensing, and cross-correlations with spectroscopic galaxy samples. Higher order statistics could also provide better constraints on $\Omega_{\rm m}$ and thus sharpen the constraints~\cite{Novaes:2024dyh,DES:2025tta}. The upcoming  DES Year 6 data release will further expand the galaxy sample and  provide  more constraining power. 

Although this work shows that the $w_0w_a$ constraints are not weakened by marginalizing over growth, the growth history still provides valuable information when constraining physical models of dark energy. For instance, constraints on modified gravity models can improve significantly when growth data are included~\cite{DES:2022ccp, DESI:2024hhd}. Recent discussions on the nature of phantom crossing~\cite{Khoury:2025txd} further emphasize the importance of growth probes. The evolution of cosmic structure remains a powerful test of dark energy physics, though a deeper understanding of astrophysical and observational systematics is essential for fully realizing its potential.

In conclusion, our results reinforce the evidence for evolving dark energy supported by the latest DESI data, and highlight the value of growth–geometry separation in improving the reliability of inference when adding weak lensing data. The framework developed in this work is broadly applicable and could be extended to other cosmological probes, such as the DESI full-shape analysis, to provide more stringent and reliable tests of dark energy dynamics.

\section*{Acknowledgements}

We thank Gary Bernstein, Dillon Brout, Josh Frieman, Dragan Huterer, Tanvi Karwal, Vivian Miranda, Shivam Pandey, Dan Scolnic and especially Martin White for useful discussions. This research used resources of the National Energy Research Scientific Computing Center (NERSC), operated under Contract No. DE-AC02-05CH11231. K.Z. and B.J. are partially supported by US Department of Energy grant DE-SC0007901.

\bibliography{ref_short}

\appendix

\onecolumngrid

\section{Impact of CMB information}\label{sec:cmb_likelihoods}


\begin{figure}
\includegraphics[width=.45\columnwidth]{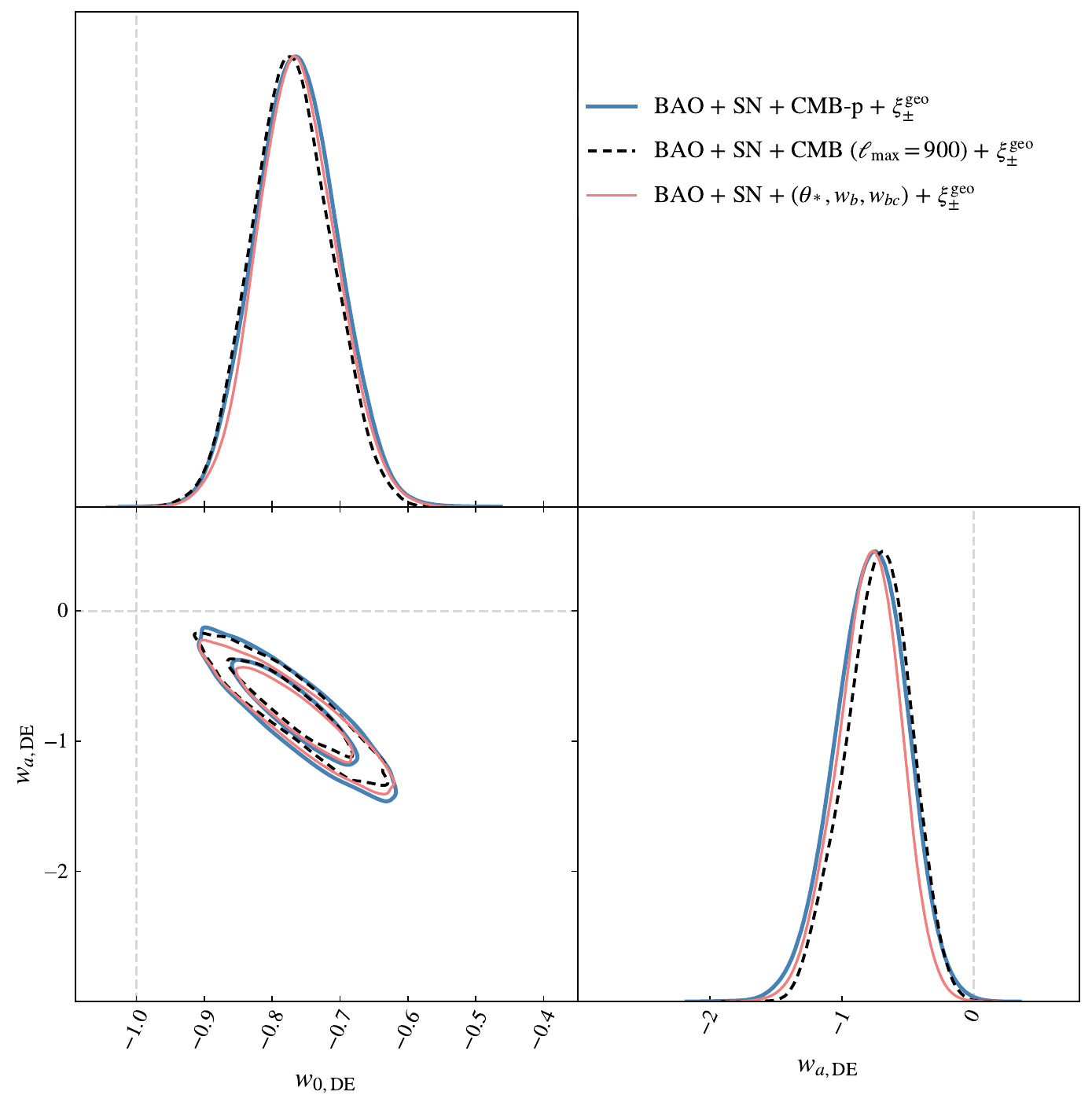}
\includegraphics[width=.45\columnwidth]{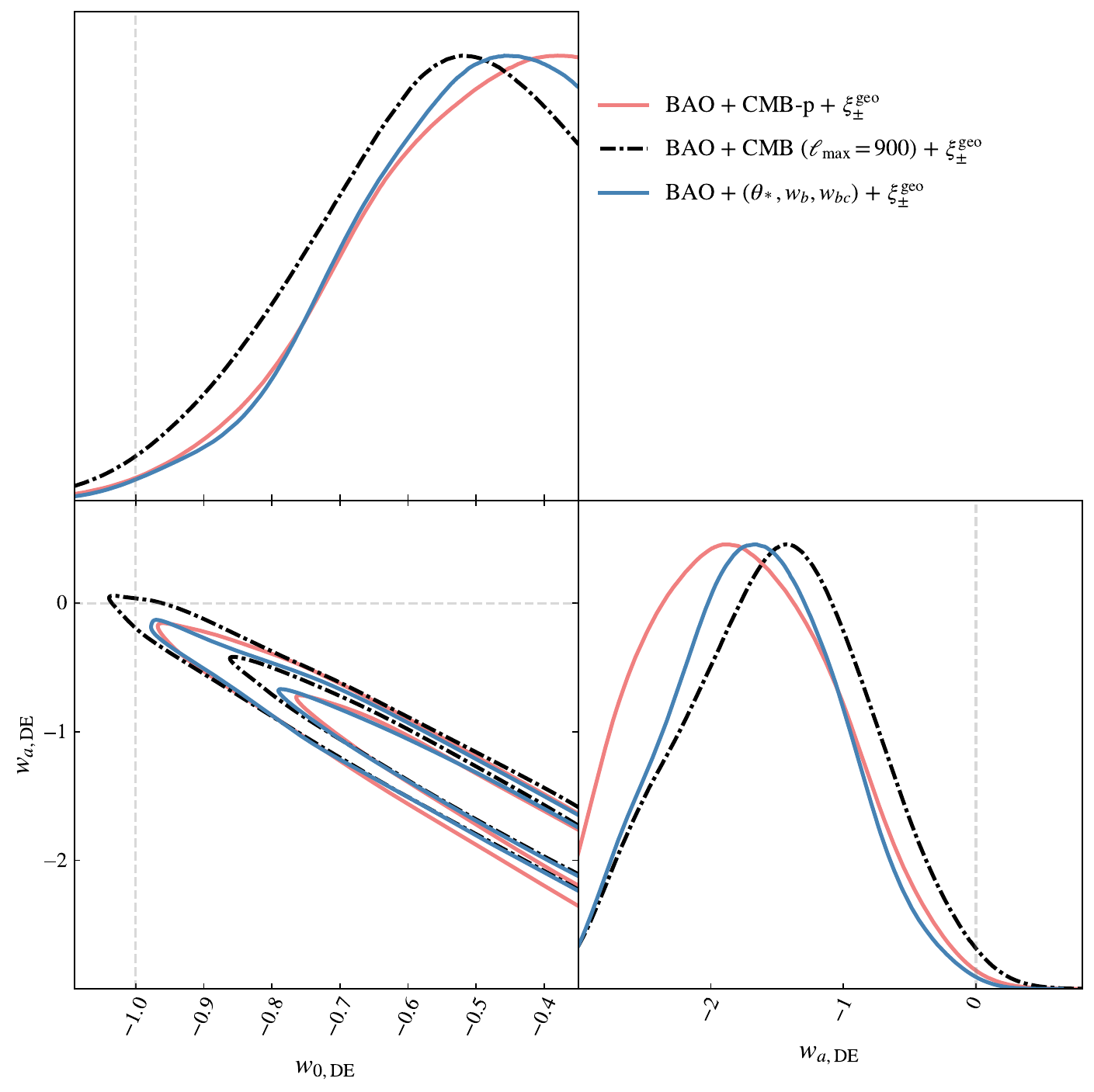}
\caption{\textit{Left:} The posterior of BAO + SN + $\xi_\pm^{\rm geo}$ and different CMB choices. The posteriors do not change much as SN dominates the constraining power. \textit{Right}: The posterior of BAO + $\xi_\pm^{\rm geo}$ and different CMB choices, without SN. Both CMB-p prior and the $(\theta_*,w_b, w_{bc})$ prior gave similar results. However, the $\ell_{\rm max}=900$ pushes the contour more toward the $\Lambda$CDM direction due to the fact that it prefers a smaller $\Omega_{\rm m}$.
}
\label{fig:cmb_comparison}
\end{figure}

In this section, we provide additional details on how different choices of CMB priors affect the results. We emphasize that in this work, CMB data are primarily treated as a prior to support inference from weak lensing probes. For this reason, we do not include the full CMB power spectrum or the CMB lensing likelihood when combining with $\xi_\pm$.

We consider three CMB likelihoods all of which include low-$\ell$ EE and high-$\ell$ TTTEEE spectra, but with varying upper cutoffs ($\ell_{\rm max}$) and marginalizations over CMB lensing. Our default choice—CMB-p—is the least informative among them in the space of $(\theta_, w_b, w_{bc})$, making it safer for combination with weak lensing data.


A similar trend was noted in Ref.\cite{DESI:2025gwf}, where replacing the full CMB likelihood with a combination of Planck ($\ell < 1000$) and ACT ($\ell > 1000$) resulted in a reduced significance for $w_0w_a$. The precise origin of this behavior remains uncertain and is beyond the scope of this paper. However, ongoing work is investigating potential systematic differences in the foreground modeling assumptions of Planck and ACT\cite{Beringue:2025bur}.

\end{document}